\documentclass[conference,10pt]{IEEEtran}
\IEEEoverridecommandlockouts
\usepackage[utf8]{inputenc}
\usepackage[T1]{fontenc}
\usepackage{cite}
\usepackage{amsmath,amssymb}
\usepackage{graphicx}
\usepackage{pifont}
\usepackage{booktabs}
\usepackage{array}
\usepackage{multirow}
\usepackage{xcolor}
\usepackage{microtype}
\usepackage{needspace}
\PassOptionsToPackage{hyphens}{url}
\usepackage[hidelinks]{hyperref}
\usepackage[capitalise,noabbrev]{cleveref}
\crefname{section}{Section}{Sections}
\crefname{figure}{Fig.}{Figs.}
\crefname{table}{Table}{Tables}
\crefname{algorithm}{Algorithm}{Algorithms}
\hypersetup{pdfauthor={Kai Ma, Quanfeng Lv, Bowei Dai, Jingguo Ge, Kefan Ruan, Ruoqi Li}, pdftitle={ScaleWire: An End-to-End Scale-Up Ethernet Simulation Framework for LLM Systems}, pdfsubject={}, pdfkeywords={}, pdfcreator={}}

\newcommand{\sys}{ScaleWire}

\definecolor{navy}{HTML}{24497F}
\definecolor{rust}{HTML}{C65F3B}
\definecolor{paper}{HTML}{F4F6FA}
\definecolor{ink2}{HTML}{52514E}
\definecolor{bufTX}{HTML}{2B4970}
\definecolor{bufReplay}{HTML}{A04536}
\definecolor{bufRX}{HTML}{5F8F8A}
\definecolor{bufReorder}{HTML}{C4A35A}

\newcommand{\BIBdecl}{\setlength{\itemsep}{0pt}}

\begin{document}

\title{\sys{}: An End-to-End Scale-Up Ethernet Simulation Framework for LLM Systems}

\author{\IEEEauthorblockN{%
Kai Ma\textsuperscript{1,2,3}\quad
Quanfeng Lv\textsuperscript{1,2,3}\quad
Bowei Dai\textsuperscript{4,*}\quad
Jingguo Ge\textsuperscript{1,2,3,*}\quad
Kefan Ruan\textsuperscript{1,2,3}\quad
Ruoqi Li\textsuperscript{1,2,3}}
\IEEEauthorblockA{%
\textsuperscript{1}State Key Laboratory of Cyberspace Security Defense\\
\textsuperscript{2}Institute of Information Engineering, Chinese Academy of Sciences\\
\textsuperscript{3}University of Chinese Academy of Sciences\\
\textsuperscript{4}Institute of Microelectronics, Chinese Academy of Sciences\\[2pt]
\href{mailto:makai@iie.ac.cn}{makai@iie.ac.cn}\quad
\href{mailto:lvquanfeng@iie.ac.cn}{lvquanfeng@iie.ac.cn}\quad
\href{mailto:daibowei@ime.ac.cn}{daibowei@ime.ac.cn}\\
\href{mailto:gejingguo@iie.ac.cn}{gejingguo@iie.ac.cn}\quad
\href{mailto:ruankefan@iie.ac.cn}{ruankefan@iie.ac.cn}\quad
\href{mailto:liruoqi@iie.ac.cn}{liruoqi@iie.ac.cn}\\
\textsuperscript{*}Corresponding authors}}

\maketitle

% Keep short columns from stretching paragraph and list spacing.
\raggedbottom

\begin{abstract}
LLM training and inference rely on scale-up networks for high-bandwidth, low-latency memory-semantic communication among accelerators (XPUs). Ethernet is emerging as an open interconnect for scale-up systems. In Ethernet systems, network-interface (NI) mechanisms across transport, network, and link layers interact, jointly shaping communication performance and on-chip resource requirements. We present ScaleWire, an end-to-end scale-up Ethernet simulation framework for LLM systems. It couples packet-level simulation of cross-layer NI mechanisms with a resource exploration engine that combines analytical estimates and simulation to size on-chip resources. We validate ScaleWire against hardware measurements and nine analytical checks, and evaluate six NI designs across four LLM communication workloads. For DeepSeek-R1 dispatch, NI design choices lead to a 4.5$\times$ difference in per-XPU on-chip storage budget. In grid-comparison experiments, the exploration engine uses 6--19\% of the simulations required by exhaustive search. Using vLLM prefill traces with kernel times measured on GH200 GPUs, ScaleWire further guides resource provisioning under a completion-time constraint for inference execution. ScaleWire thus helps designers select NI mechanisms and provision on-chip resources to balance communication performance and resource cost.
\end{abstract}


\section{Introduction}
\label{sec:intro}

Parallel training and inference of large language models (LLMs) rely on scale-up networks to provide high-bandwidth, low-latency memory-semantic communication among accelerators (XPUs)~\cite{deepseek_v3_isca,ualink_spec}.
Scale-up domains have expanded from a single server to a full rack and beyond, reaching 72 GPUs in NVL72~\cite{nvl72}, 384 NPUs in CloudMatrix384~\cite{cloudmatrix384}, and up to 1,024 accelerators in the UALink specification~\cite{ualink_spec}.
Ethernet is emerging as an open and scalable fabric for these systems~\cite{uec_spec,esun,suet_spec}.

\begin{figure}[t]
  \centering
  \includegraphics[width=0.9\columnwidth]{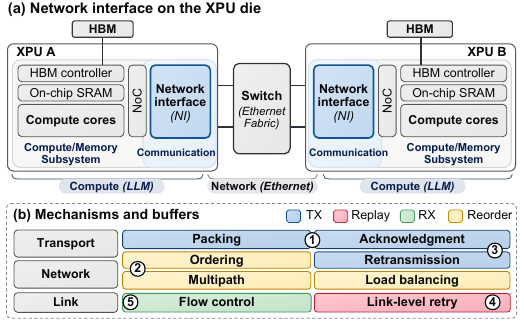}
  \caption{(a)~Network interface on the XPU die. (b)~Mechanisms in the network interface, colored by the on-chip storage resources
      they affect.
    The numbers mark five couplings, namely \ding{172}~acknowledgment and packing, \ding{173}~multipath and
      ordering, \ding{174}~acknowledgment and retransmission, \ding{175}~link-level retry and retransmission, and
      \ding{176}~hop-by-hop flow control.}
  \label{fig:arch}
\end{figure}

In an Ethernet scale-up system, the network interface (NI) bridges an XPU's compute and memory subsystem and the Ethernet fabric (\cref{fig:arch}(a)), enabling memory transactions to be exchanged among XPUs. The NI integrates transport, network, and link-layer mechanisms that consume on-chip resources such as buffers (\cref{fig:arch}(b)). NI design therefore requires selecting mechanisms and provisioning on-chip resources to balance communication performance and resource cost. This balance is difficult to achieve for two reasons.

\textbf{NI mechanisms interact across layers.}
Packetization, transmission, acknowledgment, and delivery mechanisms operate on the same packets and feedback loops. A choice in one layer therefore changes the traffic and timing observed by other layers. Acknowledgment (ACK) semantics determine whether the destination returns a separate response for each request, affecting packet formation. Multipath improves link utilization but may require receiver-side reordering. These interactions change the resources required by other mechanisms, as illustrated by the five couplings in \cref{fig:arch}(b). Prior work advances the design of these mechanisms~\cite{irn_sigcomm18,srnic_nsdi23,dixit2013spraying,vanini2017letflow,rondo2026,preface2026}.

\textbf{Resource requirements depend on the whole system.}
Even for a fixed NI design, resource requirements depend on the workload and network conditions. Data occupies on-chip storage while awaiting acknowledgments, transmission credits, or missing packets needed for in-order delivery. Workload traffic, contention, and path delays determine how long this storage remains occupied. Resource shortfalls also propagate across XPUs through backpressure, as congested receivers stall their senders. Provisioning each resource in isolation can thus violate an end-to-end performance constraint, while over-provisioning wastes die area that could serve compute and memory. Classic buffer-sizing methods for NoCs and packet networks~\cite{hu_iccad04,coenen_codes06,kumar_isvlsi11,appenzeller_sigcomm04} relate buffer capacities to traffic and performance requirements. Scale-up NI design additionally requires evaluating how choices across transport, network, and link layers change these requirements.

Existing evaluation methods address only parts of this problem. Network and LLM system simulators~\cite{ispass2013_booksim,ispass2009_garnet,ispass2023_astrasim2,simai} capture links, switches, or collective communication, but abstract away detailed NI mechanisms and their associated on-chip resource requirements. C2C-Explorer~\cite{dac2026_c2c_explorer} extends LLM-driven design-space exploration to chip-to-chip interconnects, covering packetization, scheduling, and resource-allocation choices. Although it includes buffer usage in its multi-metric evaluation, its resource metric captures relative buffer provisioning rather than the requirements of individual on-chip resources in the NI. Hardware prototypes~\cite{odcc_ethx_report} provide measurements for specific implementations, while analytical models offer fast capacity estimates for individual feedback loops. Joint modeling is therefore needed to evaluate resource requirements under the target workload. Resource capacities also form a combinatorial design space, making exhaustive simulation costly and motivating an efficient exploration method.

To address these challenges, we introduce ScaleWire, an end-to-end simulation framework for scale-up Ethernet NI design in LLM systems (\cref{fig:framework-overview}). Its packet-level simulator evaluates how NI mechanisms interact and affect communication performance and on-chip resource requirements. A resource exploration engine combines analytical estimates and simulation to provision these resources under a constraint on workload completion time.

Our contributions are threefold.
\begin{itemize}[\setlength{\topsep}{2pt}]
  \item We present ScaleWire, a packet-level simulation framework that captures interactions among transport, network, and link-layer NI mechanisms. It models their effects on completion time and on-chip resource usage. The simulator is validated against hardware measurements and supports systems with up to 1,024 XPUs.

  \item We develop a resource exploration engine that combines analytical estimates with simulation to size on-chip resources. It uses independent sizing followed by joint refinement to address performance-constraint violations when the selected resources are combined. In grid-comparison experiments, the engine requires 6--19\% of the simulations used by exhaustive search.

  \item We evaluate six NI designs across four LLM communication workloads and show how mechanism choices affect resource requirements. For DeepSeek-R1 dispatch, the per-XPU on-chip storage budget varies by up to 4.5$\times$ across designs. Using prefill traces with measured computation times, we further apply ScaleWire to provision on-chip resources under a completion-time constraint for inference execution.
\end{itemize}

\section{\sys{} Framework}
\label{sec:framework}

\subsection{Overview}
\label{sec:overview}
\begin{figure}[!htbp]
  \centering
  \includegraphics[width=0.9\columnwidth]{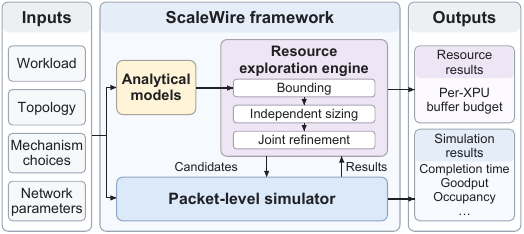}
  \caption{ScaleWire framework. Available analytical estimates guide the initial search ranges, and the engine sizes and jointly refines on-chip resources through packet-level simulation.}
  \label{fig:framework-overview}
\end{figure}

ScaleWire integrates a packet-level simulator, analytical capacity models, and a resource exploration engine (Fig.~\ref{fig:framework-overview}). Workloads are dependency graphs of communication operations and optional compute kernels with measured durations. Given a workload, network configuration, and NI design, the simulator evaluates completion time and resource occupancy. The engine uses these results to size on-chip resources under a completion-time constraint and reports a per-XPU storage budget.

\begin{figure*}[t]
  \centering
  \includegraphics[width=\textwidth,trim=1.5bp 1.3bp 4.5bp 1bp,clip]{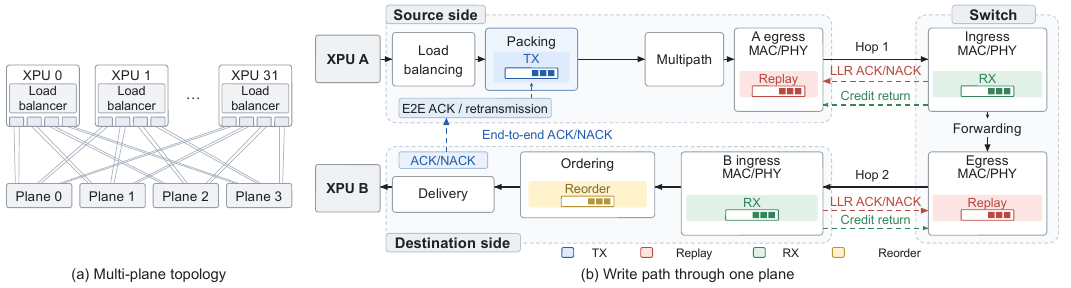}
  \caption{Scale-up topology and write path. (a) Four-plane XPU topology. (b) Processing through one plane, with the four buffer classes in color and E2E ACKs, LLR ACKs, and credit returns as dashed arrows.}
  \label{fig:overview}
\end{figure*}

\subsection{Packet-Level Simulator}
\label{sec:simulator}
Built on ns-3~\cite{ns3}, the simulator models a multi-plane Ethernet fabric in which each XPU connects to every switch plane (\cref{fig:overview}(a)). \Cref{fig:overview}(b) follows a memory write transaction (write) through one plane. At the source NI, a load balancer assigns each write to a plane. Writes are queued by destination and traffic class (TC), then packed into packets subject to payload and frame-size limits. Multipath selection chooses an egress port within the plane to transmit each packet as an Ethernet frame. At the destination, packets are checked for sequence gaps, and accepted packets are unpacked for in-order delivery to the XPU. A gap triggers a negative acknowledgment (NACK) and, with end-to-end (E2E) retransmission enabled, recovery of the missing packet.

The NI configuration determines the feedback along this path (dashed arrows in \cref{fig:overview}(b)). In delivery ACK mode, the cumulative E2E ACK advances as data is delivered in order to the destination XPU; in operational ACK mode, it advances after the write completes in destination memory. On each hop, link-level retry (LLR), when enabled, retains unacknowledged frames for retransmission after a NACK or replay timeout. Credit-based flow control (CBFC) tracks free receiver capacity as per-virtual-channel (VC) credits and permits transmission when sufficient credits are available.

Feedback and in-order delivery determine how long data stays on chip. ScaleWire tracks occupancy in four buffer classes (\cref{fig:overview}(b)). TX stores outgoing data awaiting packet formation and, with E2E retransmission enabled, retains it until acknowledged. With LLR enabled, Replay at each egress port holds frames awaiting link-level ACKs. RX at each ingress port holds incoming frames and determines available VC credits. Reorder holds out-of-order data from each source until delivery order is restored.

\subsection{Analytical Models}
\label{sec:estimator}
ScaleWire uses transmission rates and feedback delays to estimate Replay and RX capacities and, with E2E retransmission enabled, TX capacity. These estimates guide the initial search ranges.

Link-level ACKs release frames from Replay, so its capacity depends on the amount of data sent during a link round trip. Estimating this amount at the line rate and accounting for ACK spacing and the frame in transmission gives
\begin{equation}
  \widehat B_{\mathrm{Replay}}=2dr+S_A+W(k,s),
\end{equation}
where $d$ is the one-way propagation delay in ns, $r=R/8$ is the byte rate in B/ns for a line rate $R$ in Gb/s, and $S_A$ is the ACK spacing in bytes. For a packet carrying $k$ writes of $s$ bytes each, $W(k,s)=F(k,s)+20$ B is its wire occupancy. The frame size $F$ includes headers, padding, and checksums; the additional 20 B accounts for the preamble and inter-frame gap. Replay storage is measured in frame bytes, using $F$.

RX capacity determines the sender's credit window; a small window can stall transmission while credits return, even when the receiver drains data at line rate. ScaleWire models goodput over this cycle, accounting for serialization, propagation, receiver processing, unpacking, credit return, and intervening control traffic. The selected window is the smallest whose goodput is within 2\% of that with an ample window. With $g$ bytes per credit, each frame consumes $c=\lceil F(k,s)/g\rceil$ credits. Multiplying the selected credit count by $g$ gives the RX capacity estimate. This local goodput target guides the subsequent search under a workload completion-time constraint.

With E2E retransmission, TX retains writes until acknowledgment from the destination. Its estimate therefore combines three parts: data held within the switch credit windows, data injected while a forwarded frame awaits its E2E ACK, and one packet under formation. For a TX instance serving $q$ active ports, the estimate is
\begin{equation}
 \widehat B_{\mathrm{TX}}=(s+12)\left[q n_{\mathrm{sw}} k+\frac{\rho T_{\mathrm{post}}}{s}+k\right].
\end{equation}
Here, $n_{\mathrm{sw}}=\lfloor C_{\mathrm{sw}}/c\rfloor$ frames fit a switch window of $C_{\mathrm{sw}}$ credits, and $\rho=qR/8$ B/ns approximates injection using the aggregate line rate. The delay $T_{\mathrm{post}}$ spans switch egress serialization through ACK receipt at the source, including propagation, destination processing, ACK generation, and return through the switch. The three terms count writes, each occupying $s+12$ bytes including its command header.

\Cref{sec:val-models} evaluates these analytical estimates.

\subsection{Resource Exploration Engine}
\label{sec:sizing}

Exhaustive capacity search is costly, and combining independently sized resources can violate the completion-time constraint. The engine first bounds the search ranges using analytical estimates and sizes each resource independently. Joint refinement then adjusts the combination to meet the constraint.

Analytical estimates capture timing within individual feedback loops; simulation accounts for workload-dependent contention and interactions among resources. TX without E2E retransmission is sized directly through simulation, while Reorder is provisioned from observed occupancy to capture path skew and queueing.

The search varies the per-instance capacities of TX, RX, and, with LLR enabled, Replay. For a configuration $\mathbf B$, the per-XPU budget $M_{\mathrm{XPU}}(\mathbf B)$ sums these capacities over their endpoint instances and includes shared Reorder storage derived from the candidate's occupancy. Switch capacities are part of the network configuration. The engine seeks
\begin{equation}
 \min_{\mathbf B}\ M_{\mathrm{XPU}}(\mathbf B)\quad\text{subject to}\quad\eta(\mathbf B)\leq 1+\epsilon,
\end{equation}
where $\epsilon$ is the allowed fractional increase in workload completion time. To account for timing variation, the engine evaluates each candidate and the ample-buffer baseline under the same propagation-delay offsets, applied uniformly to all links. The ratio of their median completion times defines $\eta(\mathbf B)$.

\textbf{Reorder provisioning.} To capture arrival skew, the engine profiles occupancy without a Reorder capacity limit and rounds the peak over the search offsets to the allocation granularity. Candidate budgets include this provision. At held-out offsets, the engine checks whether the configuration meets the completion-time constraint and its Reorder provision covers the new peaks.

\textbf{Bounding.} Analytical estimates locate a useful starting range for each resource. The engine initially searches from half to twice the estimate, expanding within the candidate grid when the selected capacity lies at the boundary or no candidate qualifies. Resources without an estimate start from the full coarse grid.

\textbf{Independent sizing.} The engine varies one resource while keeping the others ample, selecting the smallest capacity for which that candidate and all evaluated larger capacities meet the constraint. This rule reduces sensitivity to local timing fluctuations. The engine then evaluates finer candidates around the selection and reapplies the rule, reporting failure if no capacity qualifies. The selected capacities form the starting configuration for joint refinement.

\textbf{Joint refinement.} Once the selected capacities are combined, simultaneous resource limits can increase completion time beyond the constraint. The engine evaluates candidates formed by increasing one resource class by one capacity level. It chooses the candidate with the smallest budget that meets the constraint; otherwise, it selects the lowest completion-time ratio, breaking ties by budget, and repeats. The engine stops once the constraint is met; otherwise, it reports failure when no class can grow further.

\section{Validation}

\label{sec:validation}

We validate the simulator against hardware measurements and analytical references, then assess the engine's storage budgets, constraint satisfaction, and search cost.

\subsection{Hardware Comparison and Simulation Speed}

\label{sec:val-hw}

\begin{figure}[t]
  \centering
  \includegraphics[trim=3.5bp 7.9bp 3bp 14.5bp,clip,width=0.92\columnwidth]{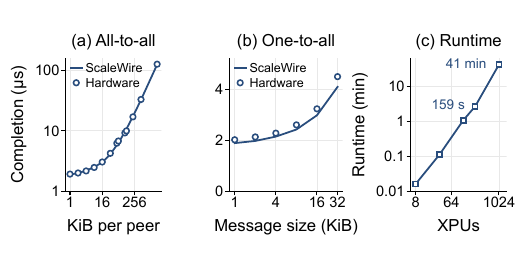}
  \caption{Hardware validation and scalability. (a) Four-host all-to-all and (b) one-to-all completion times, with one-to-all data digitized from the ETH-X report \cite{odcc_ethx_report}. (c) Single-core runtime versus XPU count.}
  \label{fig:validation}
\end{figure}

We validate ScaleWire against published measurements of the ETH-X prototype \cite{odcc_ethx_report}, which consists of four FPGA hosts connected through a 400-Gb/s switch and 5-m links. The simulation adopts the reported link and switch latencies and Ethernet encapsulation. We derive per-flit overhead from the reported byte counts and estimate NI latency from the single 128-B write measurement.

For four-host all-to-all communication across 13 message sizes (1 KiB--1.75 MiB per peer), ScaleWire achieves a mean absolute percentage error (MAPE) of 3.0\% with a 4-KiB MAC payload limit. Measured and simulated throughput both saturate near 345 Gb/s per host (Fig.~\ref{fig:validation}(a)). One-to-all MAPE is 7.3\% across six message sizes (Fig.~\ref{fig:validation}(b)).

For an all-to-all workload with 6.78 MiB per XPU, single-core runtime on an Intel Xeon Gold 6430 grows from 1.0 s at 8 XPUs to 159 s at 256 XPUs and about 41 min at 1,024 XPUs (Fig.~\ref{fig:validation}(c)). Declining packing efficiency contributes to this superlinear growth: the average number of writes per packet falls from 14.8 to 1.04 as the XPU count increases.

\Needspace{5\baselineskip}
\subsection{Analytical Validation}

\label{sec:val-models}

To check individual mechanisms, we compare the simulator with nine analytical references implemented separately. These checks cover transmission, flow control, retransmission, packing, and path selection along the write path in \cref{fig:overview}(b), as well as collectives and Poisson arrivals. The maximum deviation is 2.51\%, and six of the nine checks deviate by less than 1\%.

We assess the estimates as search starting points by comparing them with independently sized capacities for DeepSeek-R1 dispatch across 8--128 XPUs and 5-m and 50-m links. Replay and RX requirements depend mainly on link length, whereas TX demand grows with the number of destinations. Selected capacities fall within the initial bounding window of half to twice the estimate in 126 of 132 Replay/RX cases across the evaluated constraints.

\subsection{Resource Exploration Engine}
\label{sec:val-sizer}

\begin{figure}[t]
  \centering
  \setlength{\abovecaptionskip}{2pt}
  \includegraphics[trim=0bp 3bp 0bp 2bp,clip,width=\columnwidth]{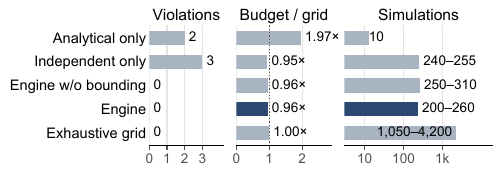}
  \caption{Engine ablation against exhaustive grid search for three workload/design cases, each under three completion-time constraints. Bars are medians and labels give ranges.}
  \label{fig:engine}
\end{figure}

We compare the engine with exhaustive grid search for DeepSeek-R1 dispatch under Default and LLR+E2E, and Qwen3 ring under Default (Fig.~\ref{fig:engine}). Bounds of 2\%, 5\%, and 10\% above each case's ample-buffer completion time give nine comparisons. Search evaluations use five propagation-delay offsets, $\{-4,-2,0,+2,+4\}$ ns; held-out checks use $\{-3,-1,+1,+3,+6\}$ ns. Reorder capacity is rounded up in 256-B units, with a 256-B minimum.

The analytical baseline uses capacity models where available and ample-buffer TX occupancy when E2E retransmission is disabled. For LLR+E2E dispatch, it violates the 2\% and 5\% constraints because the TX estimate assumes unloaded ACK timing. For Default dispatch, the occupancy-based TX estimate instead overprovisions TX storage by more than an order of magnitude. Simulation-driven independent sizing brings total budgets close to the grid optimum, but combining those capacities still violates all three constraints for Default dispatch.

In the nine grid comparisons, the engine uses 6--19\% of the simulations required by exhaustive search. Its budget matches or falls below the best grid result in eight comparisons, aided by finer capacity candidates between grid points; the remaining case requires 14\% more storage.

We further evaluate joint refinement across a broader set of workloads and designs. Across the six NI designs and four workloads of \cref{sec:case-setup}, the three constraints yield 72 configurations. Combining independently sized capacities violates the constraint in 17 cases, by up to 5.2 percentage points. Joint refinement brings all 17 within their constraints.

Across the 24 standard workload/design combinations and nine dispatch variants with skewed traffic or 128 XPUs, bounding reduces search queries by 17.27\% without changing the selected configurations.

% Declare the wide case-study figure here for the next page.
\begin{figure*}[t]
  \centering
  \setlength{\abovecaptionskip}{2pt}
  \includegraphics[width=0.95\textwidth,trim=0bp 1.7bp 0bp 0bp,clip]{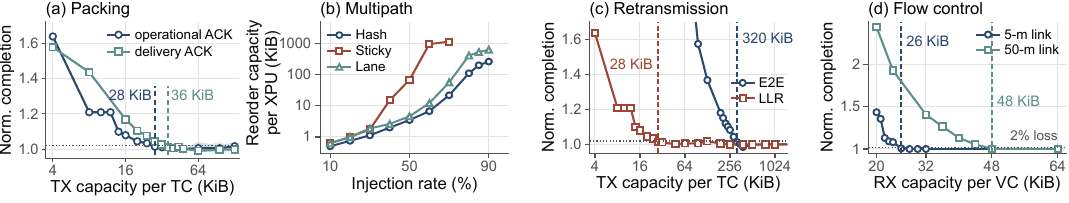}
  \caption{Effects of NI mechanisms. (a) DeepSeek-R1 dispatch completion time versus TX capacity under two ACK modes. (b) Reorder capacity under uniform random all-to-all traffic. (c) Dispatch completion time versus TX capacity under E2E and LLR. (d) Dispatch completion time versus RX capacity on 5-m and 50-m links.}
  \label{fig:mech}
\end{figure*}

At held-out offsets, median completion time meets the constraint in 67 of 72 configurations and exceeds it by at most 0.5 percentage points in the remaining five. Reorder occupancy is more sensitive to path skew, exceeding the search-based provision in 20 configurations. For provisions of at least 32 KiB, a 1.5$\times$ margin covers every held-out peak. Reorder headroom thus requires a separate check alongside completion time.

\section{NI Design Trade-offs}

\label{sec:cases}

\subsection{Setup}
\label{sec:case-setup}

We model a 32-XPU scale-up domain with four switch planes and two 400-Gb/s ports per XPU per plane, connected by 5-m links. With two TCs per NI and two VCs per port, each XPU has eight TX buffers, eight Replay buffers, and sixteen RX windows. The per-XPU storage budget sums their capacities and shared Reorder storage. Link, switch, and NI latencies follow the ETH-X setup (\cref{sec:val-hw}), and the transport mechanisms described in \cref{sec:simulator} are based on SUE-T \cite{suet_spec}. Packets carry up to 16 writes of 256 B.

Unless otherwise stated, switch receive buffers provide 64 KiB per ingress port per VC (1,024 credits at 64 B each). Switches use store-and-forward transmission and weighted round-robin scheduling. With LLR enabled, the ample-buffer configuration additionally provisions 128 KiB of switch Replay storage per port.

Default combines LLR, operational ACK, least-queued (LQ) plane selection, and per-port delivery contexts (Lane). Five variants change the recovery mechanism, ACK mode, or port-selection policy: E2E, LLR+E2E, Delivery ACK, Hash, and Sticky. LQ selects the least-loaded plane. Within a plane, Lane maintains per-port contexts, Hash pins each destination/TC to a port, and Sticky moves a context only after all its packets are acknowledged. Each design uses equal capacities for the two TCs.

Unless stated otherwise, we use error-free links and add no destination write-service delay; \cref{sec:case-e2e} examines delayed write responses. Resources are sized for a completion-time increase of at most 2\% over each design's ample-buffer baseline.

We use four write-based communication patterns from prior LLM workloads \cite{dac2026_c2c_explorer}: DeepSeek-R1 dispatch and combine \cite{deepseek_r1}, Llama-3.1-405B prefill \cite{llama3}, and Qwen3-30B training \cite{qwen3}. Each XPU sends 224 KiB to each of the other 31 XPUs in dispatch; combine doubles this volume. Llama and Qwen use rings with 62 and 248 dependent steps, respectively. Individual resource scans keep other buffers ample.

\subsection{Effects of NI Mechanisms}
\label{sec:case-mech}

NI choices change network traffic and the time data remains on chip. We examine the resulting performance and storage requirements through the five couplings in \cref{fig:arch}(b), with experiments in Fig.~\ref{fig:mech}.

\textbf{Packet formation and ACK (\ding{172}).} ACK semantics affect the response traffic generated by the destination. Delivery ACK acknowledges in-order delivery, so each write also requires an explicit response to report its completion. For DeepSeek-R1 dispatch, this raises the packet count to 3.1$\times$ that under operational ACK and extends completion time by 8\% with ample buffers. The additional traffic also raises the TX requirement from 28 to 36 KiB/TC (Fig.~\ref{fig:mech}(a)).

\textbf{Path selection and ordering (\ding{173}).} Path selection changes both link utilization and the storage needed to absorb arrival skew. Under uniform random all-to-all traffic, fixing the plane by destination limits saturation throughput to 14--25\% of link capacity. With LQ plane selection, Hash and Lane both reach 92--93\%, yet Hash requires much less Reorder storage at 80\% load: 113 versus 418 KiB per XPU. Sticky reaches 67--68\% throughput and requires nearly 1 MiB of Reorder at 60\% load (Fig.~\ref{fig:mech}(b)).

\textbf{End-to-end and link-level recovery (\ding{174}, \ding{175}).} The scope of recovery determines how long data must be retained. Even on error-free links, E2E holds data at the source until the destination ACK returns. Its TX requirement reaches 320 KiB/TC for DeepSeek-R1 dispatch and 64 KiB/TC for Qwen3 ring, compared with 28 and 6 KiB/TC under LLR alone (Fig.~\ref{fig:mech}(c)). Contention further extends retention: under all-to-all traffic, the loaded E2E ACK RTT reaches 2.3--2.7$\times$ the unloaded 1.8~\textmu{}s.

With bit errors, the shorter feedback loop of LLR also reduces recovery delay. At a bit error rate of $3\times10^{-7}$, completion time reaches 1.14$\times$ the error-free baseline for DeepSeek-R1 dispatch and 1.15$\times$ for Qwen3 ring under LLR. The corresponding ratios under E2E are 1.81$\times$ and 3.08$\times$. The ring is especially sensitive because a loss at the end of a step is detected only after the retransmission timeout, delaying the next dependent step.

\textbf{Credit flow control and backpressure (\ding{176}).} CBFC ties transmission to available receiver capacity, so endpoint RX must cover the data in flight during a credit round trip. Sustaining link utilization requires approximately 26 KiB/VC on 5-m links and 48 KiB/VC on 50-m links (Fig.~\ref{fig:mech}(d)).

Backpressure also makes the location of storage important under congestion. With four times the DeepSeek-R1 dispatch volume, a destination draining at one-quarter of the link rate raises the completion time of the other 31 XPUs to 1.98$\times$ that of a matched run without a slow receiver. Doubling switch RX to 128 KiB/VC reduces this ratio to 1.05$\times$, whereas adding the same storage to endpoint RX yields 1.92$\times$. Storage at the congested switch absorbs more of the backlog and limits its effect on other destinations.

\subsection{On-Chip Resource Requirements}
\label{sec:case-budget}

\begin{figure}[t]
  \centering
  \setlength{\abovecaptionskip}{2pt}
  \includegraphics[trim=0bp 0bp 0bp 3bp,clip,width=0.86\columnwidth]{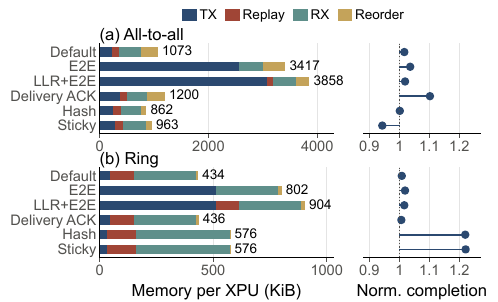}
  \caption{Per-XPU buffer budgets and completion times of six NI designs for (a) DeepSeek-R1 dispatch and (b) Qwen3 ring. Completion times are relative to Default with ample buffers, and each design is sized against its own ample baseline.}
  \label{fig:budget}
\end{figure}

\textbf{Storage requirements across NI designs.} The mechanism effects above translate into a 4.5$\times$ variation in per-XPU storage for DeepSeek-R1 dispatch (Fig.~\ref{fig:budget}). Budgets range from 862 KiB under Hash to 3,858 KiB under LLR+E2E; Default uses 1,073 KiB, with Reorder accounting for 28\%.

LLR+E2E combines source retention with per-port Replay storage (\ding{174}, \ding{175}), while Hash reduces Reorder demand by pinning each destination/TC to a port (\ding{173}). Delivery ACK's additional response traffic raises the dispatch budget by 12\% over Default (\ding{172}). Relaxing the completion-time constraint to 5\% and 10\% lowers median budgets across the four workloads by 11.5\% and 21.0\%, respectively, relative to the 2\% constraint. The dispatch budget spread remains 4.5--4.9$\times$, showing that NI choices continue to shape storage requirements under looser constraints.

\textbf{Workload-dependent design trade-offs.} Traffic distribution and communication dependencies change the relative merits of each design. For uniform DeepSeek-R1 dispatch, Hash reduces the storage budget by 20\% relative to Default's Lane policy, mainly through lower Reorder demand (102 versus 305 KiB). Under Zipf-skewed dispatch, Hash instead requires 13\% more storage than Lane.

Path selection also affects performance even with ample buffers. In that setting, Sticky completes dispatch 7\% faster than Lane, while both Hash and Sticky take about 1.18$\times$ Lane's completion time on the Llama-3.1 ring and 1.23$\times$ on the Qwen3 ring. NI selection therefore depends on both traffic distribution and communication dependencies.

\subsection{Resource Exploration for LLM Inference}
\label{sec:case-e2e}

In inference execution, compute and communication dependencies determine how network delays affect completion time. We use vLLM prefill traces for tensor-parallel Llama-3.1-70B on 8 XPUs and DeepSeek-V3.1 on 16 XPUs. Kernel durations measured on GH200 GPUs~\cite{spcl_inference_traces} are replayed along the trace dependency graph, with collectives simulated as packet-level rings on the four-plane network of \cref{sec:case-setup}. The engine sizes Default and LLR+E2E under a 2\% bound on the increase in prefill completion time.

\begin{figure}[t]
  \centering
  \setlength{\abovecaptionskip}{2pt}
  \includegraphics[trim=0bp 3bp 0bp 1bp,clip,width=\columnwidth]{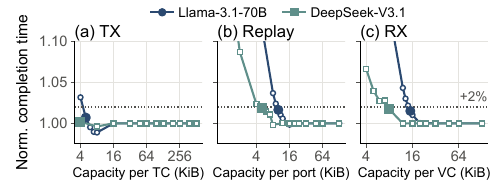}
  \caption{Resource exploration for LLM inference under the Default NI design. Points show results from independent sizing; filled points mark selected capacities, and the dotted line marks the 2\% constraint.}
  \label{fig:e2e}
\end{figure}

Fig.~\ref{fig:e2e} shows the selected capacities and where each resource begins to limit prefill completion time under Default. Across both models and NI designs, each search uses 210--240 distinct simulations, including the ample-buffer baseline.

Enabling E2E retransmission roughly doubles per-XPU storage: LLR+E2E requires 893 KiB for Llama and 502 KiB for DeepSeek, compared with Default's 421 and 267 KiB. These budgets include Reorder headroom at twice the maximum occupancy observed across the search and held-out offsets of \cref{sec:val-sizer}.

Data retained at the source until E2E acknowledgment drives this increase. Combining independently sized capacities raises completion time by 2.2\% for Llama and 2.1\% for DeepSeek. Joint refinement meets the 2\% constraint by increasing Llama's TX from 60 to 64 KiB/TC and DeepSeek's Replay from 5 to 6 KiB/port.

At five additional propagation-delay offsets, $\{-6,-5,+5,+7,+8\}$ ns, Default for Llama and LLR+E2E for DeepSeek satisfy both the completion-time and capacity checks. LLR+E2E for Llama exceeds the time constraint by 0.04 percentage points, and Default for DeepSeek exceeds its Reorder provision.

Longer destination write-response latency further increases E2E retention. Adding a 1~\textmu{}s response delay postpones operational ACKs and raises the TX requirement of LLR+E2E from 32 to 44 KiB/TC for DeepSeek and from 64 to 72 KiB/TC for Llama. Default keeps the same TX capacity. ScaleWire thus relates NI resource choices to prefill completion time while exposing the storage cost of E2E feedback.

\section{Conclusion}
\label{sec:conclusion}

This paper presents ScaleWire, an end-to-end simulation framework for scale-up Ethernet NI design in LLM systems. It combines cross-layer packet-level simulation with resource exploration to size on-chip resources under workload completion-time constraints. Across LLM communication workloads and prefill inference execution, our evaluation shows how NI mechanism interactions and workload behavior shape on-chip resource requirements and the performance--resource trade-off. ScaleWire enables designers to explore these trade-offs and jointly select NI mechanisms and resource capacities for their target workloads.


\end{document}